\documentclass[graybox]{svmult}

\usepackage{mathptmx}       
\usepackage{helvet}         
\usepackage{courier}        
\usepackage{type1cm}        
\usepackage{makeidx}         
\usepackage{graphicx}        
\usepackage{multicol}        
\usepackage[bottom]{footmisc}

\makeindex             

\begin{document}

\title*{The International X-ray Observatory and other X-ray missions,
expectations for pulsar physics}
\author{Yukikatsu Terada and Tadayasu Dotani}
\institute{Y.\ Terada \at Saitama University, 255 Simo-Ohkubo Sakura-ku Saitama city Saitama 338-8570, Japan, \email{terada@phy.saitama-u.ac.jp}
\and T.\ Dotani \at ISAS/JAXA, 3-1-1 Yoshinodai Chuuo-ku Sagamihara-city Kanagawa 252-5210, Japan, \email{dotani@astro.isas.jaxa.jp}}
%
%
\maketitle

\abstract*{Pulsar systems are very good experimental laboratories 
for the fundamental physics in extreme environments which cannot 
be achieved on ground. 
For example, the systems are under conditions of high magnetic field strength,
large gravitational potential, and fast rotation, 
containing highly-ionized hot plasmas 
with particle acceleration etc. 
We can test phenomena related to these extreme condition 
in the X-ray to sub-MeV bands. 
In future, we will get fantastic capabilities of higher sensitivities, 
larger effective area, higher energy resolutions, and
X-ray imaging capabilities with wider energy band 
than current missions, 
in addition to opening new eyes of polarization measurements, 
and deep all sky monitoring capabilities,
with future X-ray missions including ASTRO-H, eRossita, 
NuSTAR, GEMS, International X-ray Observatory (IXO) 
and so on. 
In this paper, we summarize current hot topics on pulsars 
and discuss expected developments by these future missions, especially
by ASTRO-H and IXO, based on their current design parameters.}

\abstract{Pulsar systems are very good experimental laboratories 
for the fundamental physics in extreme environments which cannot 
be achieved on ground. 
For example, the systems are under conditions of high magnetic field strength,
large gravitational potential, and fast rotation, 
representing highly-ionized hot plasmas 
with particle acceleration etc. 
We can test phenomena related to these extreme condition 
in the X-ray to sub-MeV bands. 
In future, we will get fantastic capabilities of higher sensitivities, 
larger effective area, higher energy resolutions, and
X-ray imaging capabilities with wider energy band 
than current missions, 
in addition to opening new eyes of polarization measurements, 
and deep all sky monitoring capabilities,
with future X-ray missions including {\it ASTRO-H}, {\it eRossita}, 
{\it NuSTAR}, {\it GEMS}, International X-ray Observatory ({\it IXO}) 
and so on. 
In this paper, we summarize current hot topics on pulsars 
and discuss expected developments by these future missions, especially
by {\it ASTRO-H} and {\it IXO}, based on the current design parameters.}

\section{Introduction}
\label{section:Introduction}

\begin{figure}[hbt]
\sidecaption
\centerline{\includegraphics[width=0.95\columnwidth]{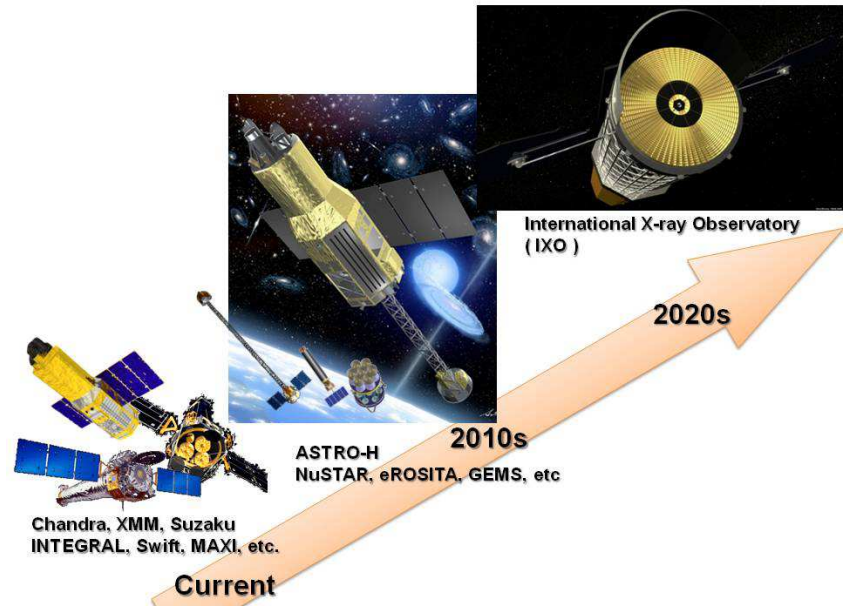}}
\caption{Artists view of the current and next-generation X-ray missions}
\label{fig:future_missions}
\end{figure}

Currently, we have nice missions like {\it Chandra}, {\it Newton}, and 
{\it Suzaku}, as well as {\it INTEGRAL}, {\it Swift}, {\it MAXI}, and so on.
In very near future in 2010s, we will have the {\it ASTRO-H} mission 
in addition to small satellites like {\it NuSTAR, eROSITA, GEMS}, etc.
After the decade, we will have an ultimate X-ray mission, 
named international X-ray observatory ({\it IXO}) in 2020s
as shown in Fig.\ref{fig:future_missions}.
The performances of these future missions are summarized 
in section \ref{section:future_missions}.

\begin{table}[hb]
\caption{Key Science on future missions.}
\label{tab:keyscience}
\begin{tabular}{p{2cm}p{3.1cm}p{3.1cm}p{3.1cm}}
\hline\noalign{\smallskip}
Objects$^a$ & {\it IXO} (2021--) & {\it ASTRO-H} (2014--) 
            & {\it NuSTAR} (2011--)\\
\noalign{\smallskip}\svhline\noalign{\smallskip}
BH,NS,AGN & $\bullet$ {\bf Matter under Extreme Condition}
          & $\bullet$ {\bf Physics under extreme environment}
          & $\bullet$ Survey of Massive BH \\
{}        &
          & $\bullet$ Relativistic space-time near BH
          & $\bullet$ Galactic Center Region \\
{}        &&&\\
SNR,SN,Star, BH,AGN 
          & $\bullet$ Life cycles of matter and energy
          & $\bullet$ Evolution of Cluster of galaxies
          & $\bullet$ Nucleaosynthesis in SN \\
{}        &
          & $\bullet$ Co-evolution of BH and galaxy
          & \\
{}        &&&\\
SNR,NS, WD,CG
          &
          & $\bullet$ {\bf Variety of the non-thermal universe}
          & $\bullet$ {\bf Cosmic-ray accelerator} \\
{}        &&&\\
CG        & $\bullet$ Formation of Structure
          & $\bullet$ Evolution of Cosmic structure with DM
          & \\
\noalign{\smallskip}\hline\noalign{\smallskip}
\end{tabular}
$^a$ BH: black hole, NS: neutron star, WD: white dwarfs,
SNR: supernova remnant, SN: supernova, 
AGN: active galactic nuclei, CG: cluster of galaxies
\end{table}

 Then, what should we do on pulsars with future missions?
Table~\ref{tab:keyscience} summarizes the key sciences of 
{\it NuSTAR}, {\it ASTRO-H}, and {\it IXO}, 
which are officially announced. 
These key topics cover many sciences widely, 
but they can be roughly categorized into four classes; 
(1) the fundamental physics on extreme environment, 
(2) the life cycle of materials, 
(3) non thermal emissions from accelerated particles, 
and 
(4) the formation of cosmic structure. 
Among these items, topics related to neutron stars are shown 
in bold style in the table. Therefore, one of our goals on pulsars 
with future missions is in the fundamental physics on extreme 
environments. In this paper, we pick up the following items
and describe in detail in section \ref{section:science},
as examples of pulsar sciences with future missions.

\medskip
\begin{itemize}
\item General relativity under strong gravity,
\item Equation of State in neutron stars,
\item Plasma physics under a strong magnetic field,
\item Emission mechanism from Magnetars,
\item Diversity of Pulsar systems, white dwarf pulsars.
\end{itemize}

\section{Future X-ray Missions}
\label{section:future_missions}

\subsection{Overview of X-ray Missions}
\label{section:mission:overview}

\begin{figure}[ht]
\sidecaption
\includegraphics[width=0.5\columnwidth]{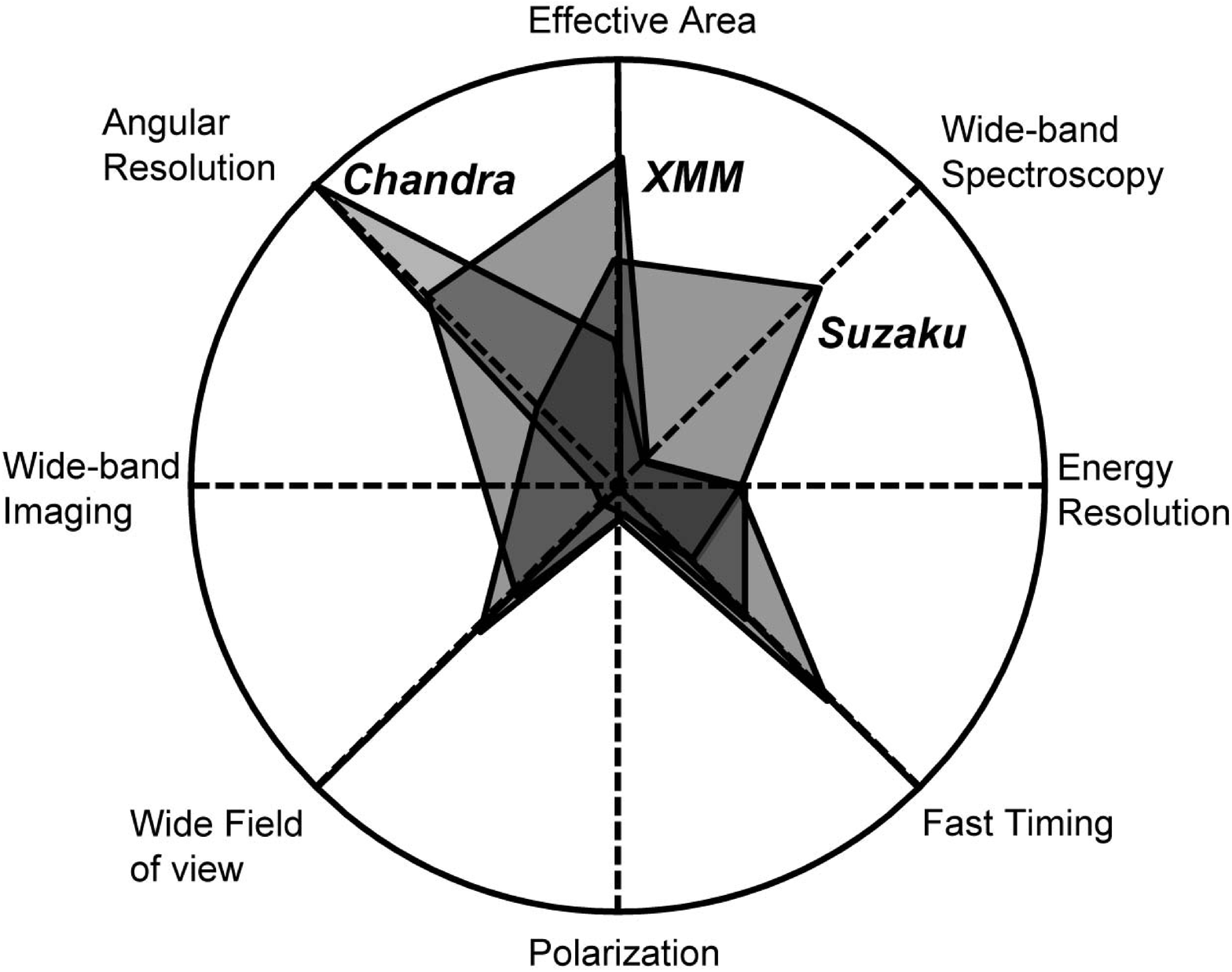}
\includegraphics[width=0.5\columnwidth]{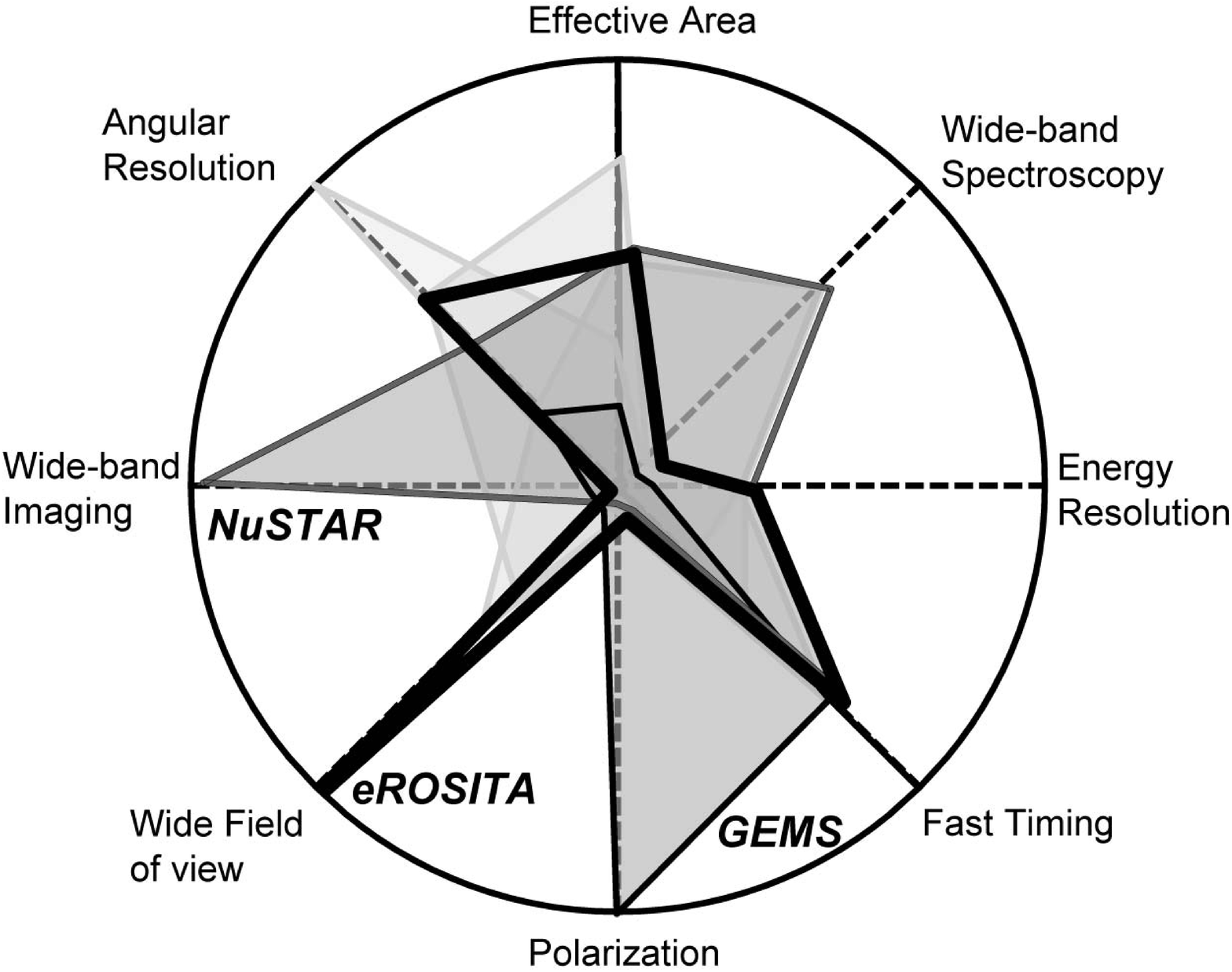}
\includegraphics[width=0.5\columnwidth]{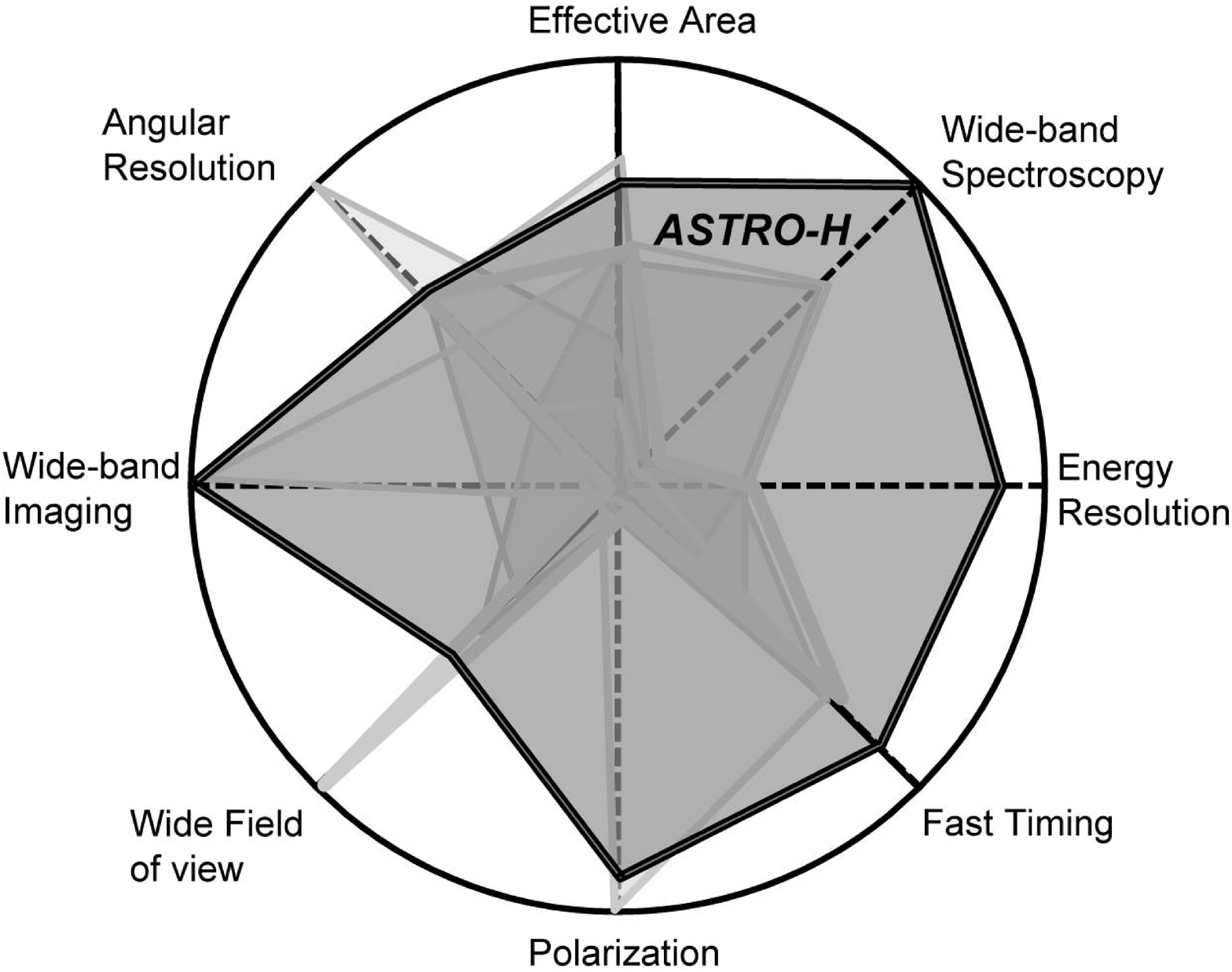}
\includegraphics[width=0.5\columnwidth]{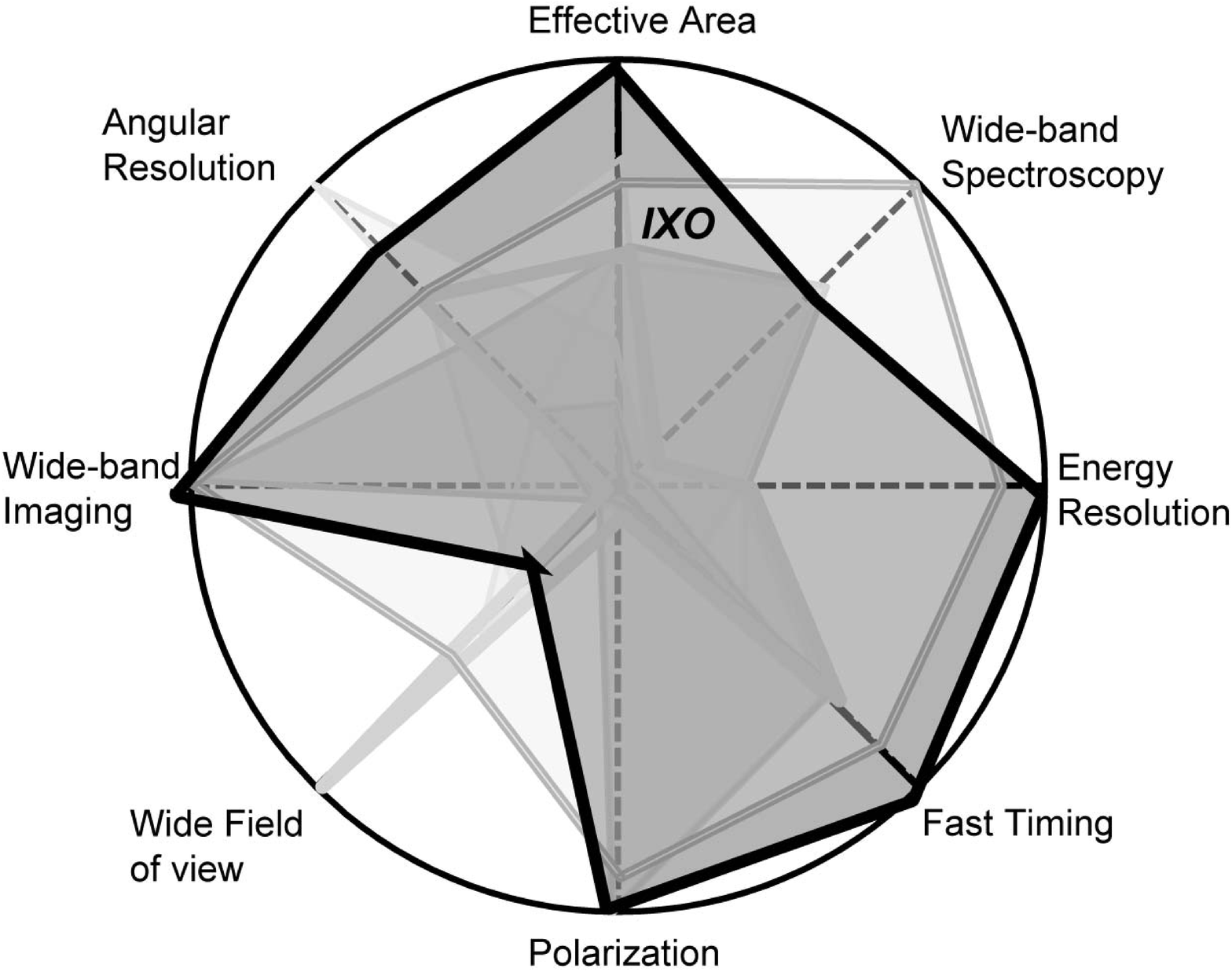}
\caption{Comparison of features between X-ray missions. 
(Top left) diagram of the current missions, {\it Suzaku}, 
{\it XMM-Newton}, and {\it Chandra}.
(Top right) the next small missions, {\it NuSTAR}, 
{\it eROSITA}, and {\it GEMS}.
(Bottom left and right) {\it ASTRO-H} mission and the {\it IXO}.}
\label{fig:missions}
\end{figure}

Many kinds of X-ray missions are listed 
in section \ref{section:Introduction},
such as {\it Suzaku, Chandra, XMM-Newton, 
NuSTAR, GEMS, eROSITA, ASTRO-H}, and {\it IXO}. 
Specific features and characteristics of these missions 
can be summarized in diagrams of Fig.~\ref{fig:missions}.
Imaging capabilities, like effective areas, angular resolutions, 
wide-energy band-pass, and the field of view, are indicated in the 
left 1/3 hemisphere of the diagram. 
Spectroscopic features, 
like energy resolutions and wide-band sensitivities, 
are shown in the right 1/3 part. 
Other additional characteristics, 
like fast timing capabilities and polarization measurement,
are plotted in the bottom part of the diagram.

As shown in the Fig.~\ref{fig:missions} top left, 
the current missions have complementary performances; 
{\it Chandra} has super high angular resolution, 
{\it XMM-Newton} has large effective area, 
and {\it Suzaku} has wide-band and high-sensitive 
spectroscopic performances.  
Thus, these three current missions covers 
upper hemisphere of the diagram, complimentary.

In very near future, 
we will have small satellite missions, 
named {\it NuSTAR, GEMS}, and {\it eROSITA}. 
As shown in the Fig.~\ref{fig:missions} top right, 
these three missions will explore new capabilities 
in the left bottom part of the diagram; 
i.e., hard X-ray imaging, polarization, 
and deep all sky survey. 
The details are summarized in section \ref{subsection:mission:smex}.

Following these small missions, 
we will have general-purpose X-ray observatories, 
{\it ASTRO-H} and {\it IXO}, 
covering high performances in the diagram, 
as shown in Fig.~\ref{fig:missions} bottom.
Most prominent feature of {\it ASTRO-H} is 
the high energy-resolution spectroscopic feature 
with the micro-calorimeter array. 
{\it IXO} will have an excellent X-ray mirror 
with huge effective area of 30,000 cm$^2$.
The detail descriptions of {\it ASTRO-H} and {\it IXO} 
are given in sections \ref{subsection:mission:astroh} 
and \ref{subsection:mission:ixo}, respectively.

\medskip

In summary, future missions will have 
capabilities of the followings.
\begin{itemize}
\item High energy-resolution spectroscopy({\it ASTRO-H}, {\it IXO})
\item Hard X-ray imaging ({\it NuSTAR}, {\it ASTRO-H})
\item Huge collection area of X-rays ({\it IXO})
\item High sensitivities in soft gamma-ray band ({\it ASTRO-H})
\item All sky monitoring function ({\it eROSITA})
\item Polarization measurement ({\it GEMS}, {\it ASTRO-H})
\end{itemize}

\subsection{The Small Satellite Missions in 2010s}
\label{subsection:mission:smex}

\begin{figure}[htb]
\sidecaption
\centerline{\includegraphics[width=0.8\columnwidth]{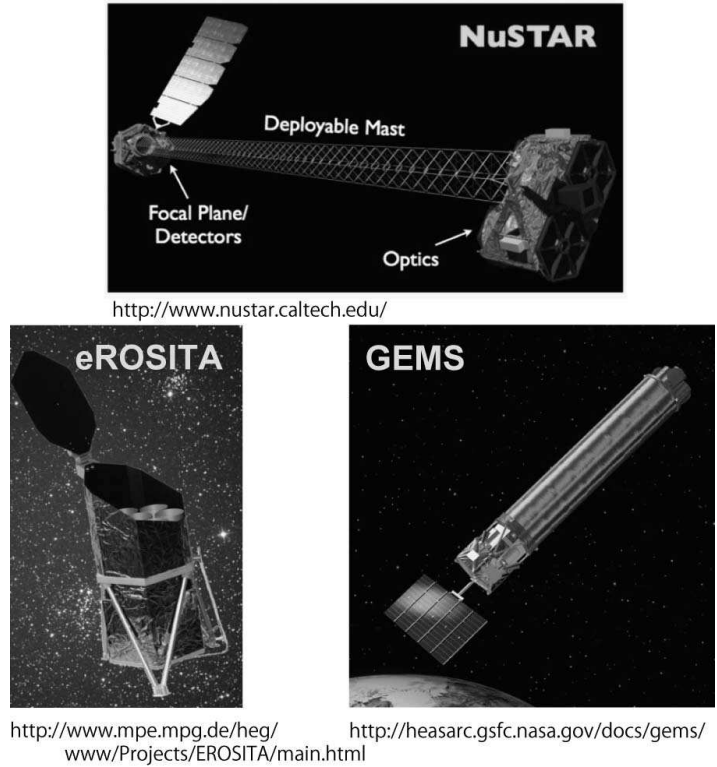}}
\caption{Artists view of the next-generation small missions, 
{\it NuSTAR}, {\it eROSITA}, and {\it GEMS}.}
\label{fig:smex}
\end{figure}

The nuclear spectroscopic telescope array ({\it NuSTAR}) is 
a NASA SMEX (small explorer) mission, 
which is planned to be launched in 2011 
to archive the hard X-ray imaging in 5 to 80~keV band
for the first time, carrying the focusing hard X-ray telescope.
The previous X-ray mirrors utilize total-reflection 
to focus soft X-ray photons.
It requires very small incidence angles, and therefore, in principle,
the reflection energies with the practical design of mirrors 
are limited in the soft X-ray band below 10 keV\@.
The new hard X-ray mirrors use the Bragg reflection process 
to focus hard X-ray photons. 
The mirrors are coated by multi-layers to reflect hard X-rays 
via Bragg reflection to cover up to 80 keV band 
with the effective area of 70 -- 1000 cm$^2$.
The focal length of the optics is as long as 10 meters
as illustrated in Fig.~\ref{fig:smex}.
The current design of {\it NuSTAR} represents 
the angular resolution of 7.5 arcsec (FWHM) and 
energy resolution of 1.2 keV at 68 keV with CdTe focal-plane detectors.

The extended ROentgen Survey with an Imaging Telescope Array 
({\it eROSITA}) will perform the first imaging all-sky survey 
in the medium energy X-ray range up to 10 keV 
with an unprecedented spectral and angular resolution. 
The instruments will be carried by Spectrum-Roentgen-Gamma (SRG) satellite, 
which will be launched in 2012. 
The telescope will consist of seven X-ray optics, 
which consist of mirrors and pn-CCD cameras developed for XMM-Newton. 
The on-axis effective area will reach 2000~cm$^2$ 
with seven telescopes in total. 
The satellite will be operated at three types of 
all-sky scanning observation modes, 
covering 20,000 deg$^2$ in 3 years with the sensitivity 
at the $10^{-14}$ erg cm$^{-2}$ s$^{-1}$ level below 2 keV band, 
or 200 -- 300 deg$^2$ at $10^{-15}$ erg cm$^{-2}$ s$^{-1}$ 
sensitivity below 2 keV band.

The Gravity and Extreme Magnetism SMEX ({\it GEMS}) is also 
a NASA SMEX mission, which will be launched in 2014. 
The mission performs X-ray polarization measurements 
in the 2 -- 10 keV band with the photo-electron tracking 
gas-pixel detector. 
The energy range is mainly determined by the X-ray telescope,
which covers below 10 keV band having 
a focal length of 4.5 meters as illustrated in Fig.~\ref{fig:smex}.
The goal of the current design of the polarization measurement is 
the minimum detectable polarization (MDP) at 0.2\% (99\% confidence)
at 1 Crab flux or MDP of 2\% at 10m Crab level \cite{gems}.

\subsection{The {\it ASTRO-H} Mission}
\label{subsection:mission:astroh}

The {\it ASTRO-H} mission is the sixth series of the Japanese X-ray satellites
\cite{next08}, which will be launched in 2014. 
This is the Japan and US collaboration mission with a part of ESA members.
In addition to engineering interests on many challenging instruments 
and the spacecraft bus system using SpaceWire networks 
as a future standard satellite, 
{\it ASTRO-H} will be a multi-purpose observatory of many scientific topics 
as already listed in Table \ref{tab:keyscience}.

\begin{table}[h]
\caption{Design parameters of {\it ASTRO-H}.}
\label{tab:astroh}
\begin{tabular}{p{3cm}p{3cm}p{5.5cm}}
\hline\noalign{\smallskip}
Optics    & Instruments & Parameters$^a$ \\
\noalign{\smallskip}\svhline\noalign{\smallskip}
SXS + SXT       & Micro Calorimeter   & $\Delta$ E = 7 eV (4 eV goal)\\
(0.3 -- 10 keV) & + Soft X-ray Mirror & EA = 260 cm$^2$ at 6 keV\\
                & &\\
SXI + SXT       & P-ch CCD camera     & $\Delta$ E = 150 eV \\
(0.4 -- 10 keV) & + Soft X-ray Mirror & FOV = 38 arcmin\\
                &                     & PSF = 1.7 arcmin(FWHM)\\
                &                     & EA $\sim$ 368 cm$^2$ at 6 keV\\
                & &\\
HXI + HXT       & DSSD+CdTe stack     & $\Delta$ E $\sim$ 2 keV \\
(5 -- 70 keV)   & surrounded by BGO   & FOV = 9 arcmin\\
                & + Hard X-ray Mirror & PSF = 1.7 arcmin(FWHM)\\
                &                     & EA $\sim$ 300 cm$^2$ at 30 keV\\
                &                     & 1 -- 10 $\mu$ Crab sensitivity$^b$ \\
                & &\\
SGD             & DSSD+CdTe stack     & $\Delta$ E $\sim$ 2 keV \\
(10 -- 600 keV) & surrounded by BGO   & FOV = 33.3 arcmin\\
                & (no focusing mirror)& EA $\sim$ 100 cm$^2$\\
                &                     & 0.1 -- 1 m Crab sensitivity$^b$\\
\noalign{\smallskip}\hline\noalign{\smallskip}
\end{tabular}
$^a$ $\Delta$ E means the energy resolution, EA means the effective area,
PSF means the point spread function (or angular resolution), 
and FOV means the field of view. \\
$^b$ Sensitivity at 1 M sec exposure.
\end{table}

 {\it ASTRO-H} carries four types of X-ray optics, 
SXS+SXT (Soft X-ray Spectrometer plus Soft X-ray Telescope), 
SXI+SXT (Soft X-ray Imager plus Soft X-ray Telescope), 
HXI+HXT (Hard X-ray Imager plus Hard X-ray Telescope), 
and SGD (Soft Gamma-ray Detector), as listed in Table \ref{tab:astroh}.
The design parameters of these instruments 
are also summarized in the table. 

The most important feature of {\it ASTRO-H} is 
the high resolution spectroscopy with the micro calorimeter array, 
named SXS, with the energy resolution of about 7 eV (or 4 eV as a goal)
with a rather large effective area of 200 cm$^2$ \cite{sxs08}.
In ground developments, they have already achieved 
the energy resolution of 3.8 eV. 

 The second importance is the hard X-ray imaging capability 
up to 80 keV band by the HXI+HXT system 
with the larger effective area of 300 cm$^2$ 
at 30 keV than that of {\it NuSTAR}.
The HXT employs the ``multi-layer super mirror'' 
technique using the Brag reflection 
as described in section \ref{subsection:mission:smex}, 
which already verified by balloon experiments of {\it InFOC$\mu$S} series.
The focal plane detector of the HXT is 
the hard X-ray instrument named HXI, 
which is the stack of 4 DSSDs (double sided Si strip detectors) 
and 1 CdTe strip detector surrounded by the BGO crystals 
as anti-coincidence shield counters \cite{hxi08}.
The mission also carries the CCD camera, named SXI, 
on the focal plane of the other SXT,
with the wide field of view of about 38 arc minutes
in the 0.4 -- 10 keV band.

The third importance of this mission is 
the super high sensitivity in the soft gamma-ray energy band. 
This feature is achieved by the SGD (soft gamma-ray detector), 
which is Si-and-CdTe Compton-cameras with a narrow field of view.
Although the SGD employs the Compton camera technique, 
the detector has little imaging capability, 
achieving quite high sensitivities at 1 mCrab level 
at the sub-MeV energy range 
by using the Compton kinematics as a powerful method 
to reject backgrounds \cite{sgd05,sgd09}.
In other words, the {\it ASTRO-H} will improve their sensitivities 
in the hard X-ray band above 10 keV band by one or two order of magnitude 
by the HXI and the SGD, respectively,
than the current mission.

\subsection{The International X-ray Observatory}
\label{subsection:mission:ixo}

The international X-ray observatory ({\it IXO}) is 
an international mission in collaboration with 
NASA, ESA, and JAXA to be realized in 2020s \cite{ixo10}. 
The satellite is now planned to be launched as early as 2021, 
combining a super large X-ray mirror with many instruments 
as listed in Table \ref{tab:ixo}.

\begin{table}
\caption{Design requirements for {\it IXO} instruments.}
\label{tab:ixo}
\begin{tabular}{p{3cm}p{3cm}p{5.5cm}}
\hline\noalign{\smallskip}
Instruments     & Detector       & Parameters \\
\noalign{\smallskip}\svhline\noalign{\smallskip}
Mirror          & soft X-ray Mirror &  EA = 3 m$^2$ at 1.25 keV   \\
                & soft X-ray Mirror &  EA = 0.65 m$^2$ at 6 keV   \\
                & hard X-ray Mirror &  EA = 0.015 m$^2$ at 30 keV \\
                &                   & \\
XMS             & Micro Calorimeter & $\Delta$ E = 2.5 eV (FWHM) and FOV = 2 arcmin \\
 (0.3 -- 7 keV) &                   & $\Delta$ E = 10 eV (FWHM) and FOV = 10 arcmin \\
                &                   & PSF = 5 arcsec (HPD) \\
                &                   & \\
WFI/SXI         & CCD camera        & $\Delta$ E = 150 eV (FWHM) and FOV = 18 arcmin \\
 (0.1 -- 15 keV)&                   & PSF = 5 arcsec (HPD) \\
                &                   & \\
WFI/HXI         & Si+CdTe+BGO       & PSF = 30 arcsec (HPD) \\
 (7 -- 40 keV)  &                   & \\
                &                   & \\
XGS             & Grating Optics    & E/$\Delta$ E = 3000 \\
 (0.3 -- 1 keV) &                   & \\
                &                   & \\
HTRS            & Si Drift          & $10^6$ cps with $<10$\% deadtime\\
                &                   & \\
XPOL            & Polarimetry       & 1\% MDP, 100 ksec, $5\times 10^{-12}$ cgs (2-6 keV)\\
\noalign{\smallskip}\hline\noalign{\smallskip}
\end{tabular}
\end{table}

The most prominent feature of {\it IXO} is 
a huge single X-ray mirror with a 3.0 m$^2$ collecting area 
and 5 arc-second angular-resolutions in the softer energy band. 
The energy coverage of the X-ray mirror will be extended to 
the hard X-ray range with an effective area of 150 cm$^2$ at 30 keV, 
with a focal length of about 20 meters.

{\it IXO} will carry a micro calorimeter detector, 
named XMS (X-ray Micro calorimeter Spectrometer), 
which is the next generation of the micro-calorimeter array 
from {\it ASTRO-H}, adopting the transition-edge-sensor technique. 
The XMS will achieve the energy resolution of about 2.5 eV 
in the 0.3 -- 7 keV band.  {\it IXO} also carries a grating optics, 
named XGS (X-ray Grating Spectrometer), 
achieving a resolving power of $E/\Delta E \sim 3000$.

{\it IXO} also have a wide band imager, 
a kind of a hybrid detector consisting of WFI (Wide Field Imager) 
and HXI (Hard X-ray Imager), 
which have almost the same design as the SXI and HXI on-board {\it ASTRO-H}.
The mission also has new types of instruments, 
such as HTRS (High Time Resolution Spectrometer) and 
XPOL (X-ray Polarimeter), as listed in Table \ref{tab:ixo}.
The HTRS is a silicon drift sensor having a high count-rate acceptance 
up to $10^6$ counts-per-second with less than 10 \% dead time. 
The XPOL is a a fine-grid Gas-Pixel-Detector 
having high polarization sensitivity 
reaching 1 \% level for a 1mCrab full-polarized light 
with 10 ksec exposure.

\section{Pulsar Sciences with Future X-ray Missions}
\label{section:science}

\subsection{General relativity under strong gravity}
\label{subsection:science:gr}
X-rays from compact accreting objects come from in-flowing materials 
onto the objects. Therefore, we can probe 
the extreme environment near the compact object
via X-ray spectroscopy, such as strong gravity 
and relativistic motion of the accreting matter. 

\subsubsection{Recent Studies}
\label{subsection:science:gr:recent}

The discovery of the gravitationally red-shifted line emission 
from a massive black hole, active galactic nuclei MCG-6-30-15, 
is one of the most important results with {\it ASCA} \cite{tanaka95}. 
They found a red-shifted double-wing line structure of Fe K line 
in the X-ray spectrum taken with {\it ASCA}. 
Such a profile was theoretically predicted by Fabian et al (1984)
six years before the observational report \cite{fabian84}. 
The profile is caused by a Doppler effect of the rotating motion 
on the accretion disk as well as the gravitational red shift 
by the strong gravity of the black hole 
following the Einstein's general theory of relativity.

Recently, gravitationally red-shifted line emissions are reported 
from several neutron star systems, such as Serpens X-1 \cite{Bhattacharyya07}, 
4U1820-30, and GX349+2 \cite{Cackett08} with {\it XMM-Newton} and {\it Suzaku}.
However, the line profile has a very extended low-energy wing, 
and it is difficult to distinguish from continuum emissions 
when the observation is in a narrow energy range. 
Thus, some other groups pose a question on the discovery of the 
gravitational red-shifted emissions \cite{yamada09}.
They argue that if they take into account the pile-up effects of 
the detector correctly, then the wide wings around the Fe line band 
are disappeared; i.e., the spectra do not require the gravitationally 
red-shifted profile. 
This situation is mainly caused by the difference of the definition 
of the continuum model. 
So, the wide-band spectroscopy is the key of this observation. 

\subsubsection{Future prospects}
\label{subsection:science:gr:future}

The recent argumentation on the discovery of gravitationally 
red-shifted emissions from neutron stars is mainly caused by 
the difference of the continuum models. 
In future works, we first have to verify of the line profiles 
with wide band spectroscopy in the X-ray band.
{\it ASTRO-H} will detect X-ray emissions of objects up to 400 keV 
with very high sensitivities in the hard X-rays to the soft gamma-ray band.
Then, we can test a complex continuum model 
by separating several components, 
such as reflection radiation, Compton-scattered component, 
disk black body emission, etc, to recognize the line structure.

Then, we can enjoy the general relativity around neutron stars.
With the huge effective area of the {\it IXO} mission, 
it can catch line profiles in a very fast time resolution
with very good statistics. 
According to the calculation of the time evolution of X-ray spectra 
from an accretion disk \cite{Armitage03}, 
we can catch the trajectory of accreting blobs 
one by one \cite{Brenneman09}.
Then, as well as the Doppler effect of the rotation of the accretion disk,
the gravitational red-shifts as a function of in-falling time 
will be observed directly with the {\it IXO} \cite{Brenneman10}. 
Therefore, this study will directly link 
to the challenges of the verification of the general relativity.

\subsection{Equation of State in neutron stars}
\label{subsection:science:eos}
Inner structure of a neutron star is highly dependent on 
the physics in high energies \cite{Lattimer07}. 
Currently, it is considered to have a shell structure, 
from center to surface: 
inner core, outer core, inner crust, outer crust, and atmosphere. 
This structure strongly depends on the equation of state of 
cold materials of neutron stars 
in high-density and high-pressure environments under a strong gravity. 

\subsubsection{Recent Studies}
\label{subsection:science:eos:recent}

The equation of states of matters in a neutron star, 
described by the pressure and the density, 
can be observed by global parameters 
like mass and radius of the object.
The mass-radius relationships for neutron stars, 
reflecting the equation of states of cold super-dense matter, 
are calculated in several conditions by many authors \cite{Lattimer07}. 
Thus, measurements of the mass and the radius
are important for the tests of these theoretical expectations.

Observationally, we can measure the mass of a neutron star 
from the binary motion if the object has a companion star
\cite{Lattimer07}, but it is limited to very few objects. 
Radius of a neutron star can be derived 
by the flux of the black body emission with the Stefan-Boltzmann's low
when the distance to the object is known, 
but there are many difficulties in the infrared and optical observations
of black body component,
because of many contamination lights \cite{Kaspi06}.  

Measurement of gravitationally red-shifted emission or absorption lines
will give us the information of gravity; 
i.e., the ratio between the mass and the radius. 
This method is already successful in the white dwarf cases \cite{Provencal98}.
In 1980s, absorption feature was found with {\it Tenma} and {\it EXOSAT}
from neutron stars, X1636-536, X1608-52 etc, 
during its X-ray bursts \cite{waki84,nakamura88,Langmeier86}, and 
studies of estimation of the mass and radius were performed \cite{fujimoto86}.
Recently, a red-shifted absorption lines are found from a neutron star, 
EXO0748-676, during the X-ray flare, 
in the grating spectra of XMM-Newton satellite \cite{Cottam02}. 
They found many significant absorption features with a red-shift of z=0.35, 
which is consistent with models of neutron stars 
if they assume the mass of the neutron star is 
in a range of typical value of 1.3 to 2.0 solar masses. 

\subsubsection{Future prospects}
\label{subsection:science:eos:future}

With the current instruments, 
from absorption lines on X-ray spectra of an atmosphere on a neutron star, 
we measure only the gravitational red-shift value, 
as an observational parameter 
related to the mass and radius of the neutron star. 
In other words, we cannot obtain the mass and radius 'independently'
with the current instruments. 

In future, the {\it IXO} provides the sensitive X-ray spectroscopy and 
fast photometry with high energy resolutions, 
and thus {\it IXO} can achieve phase-resolved high-resolution spectroscopy. 
The spectroscopic observation of X-ray bursts with {\it IXO}, 
showing the atomic absorption lines, will give us information 
of the acceleration of gravity at the stellar surface and 
the gravitational red-shift separately \cite{Ozel03}, 
through the X-ray spectroscopic features 
by the pressure broadening 
and general relativistic effects \cite{Bhattacharyya06}. 
Therefore, we can determine the radius and mass separately 
with the {\it IXO} \cite{Paerels10}. 
In this way, it is feasible to test the equation of states of neutron stars 
by measuring the mass and the radius of the objects with the future missions. 

\subsection{Plasma physics under a strong magnetic field}
\label{subsection:science:plasma}
Accreting binary pulsars are good targets 
to study radiation transfer of X-ray photons 
under strong magnetic field in a range of $10^{12}$ Gauss. 
Several objects show the cyclotron resonance scattering features (CRSFs) 
in their X-ray spectra, 
because the cyclotron resonance energy  $E_{\rm a}$ comes 
in the X-ray energy band as $E_{\rm a} = 11.6 B_{12} (1+z_{\rm g})^{-1}$ keV, 
where $B_{12}$ is the magnetic field strength in unit of $10^{12}$ Gauss 
and $z_{\rm g}$ is the gravitational red-shift at the resonance points. 
Since the first discovery of CRSF from neutron stars from Her X-1
\cite{Trumper78}, the cyclotron resonance features are well studied 
with many X-ray missions on about 15 X-ray pulsars \cite{makishima99}. 
However, there still remain many mysteries 
on the radiation transfer process under the strong magnetic field.

\subsubsection{Recent Studies}
\label{subsubsection:science:plasma:recent}

Observation of CRSF in the X-ray band is one of the direct measurements 
of the magnetic field strength of neutron stars. 
In this measurement, the resonance energy $E_{\rm a}$ is considered 
to represent the surface magnetic field strength of the object. 
Interestingly, 
recently, several X-ray pulsars such as 4U 0115+63 and X0331+53 
show the changes of $E_{\rm a}$ 
as a function of their X-ray luminosities \cite{nakajima06};
the cyclotron energies are increased when X-ray luminosity became low. 
This phenomenon should be due to the variation of the height 
of the X-ray emitting accretion column 
by their luminosity or accretion rate.
This means that, if we observe the object in a lower luminosity phase, 
we will measure the magnetic field at lower part of the column 
near the surface of the neutron star. 
Nakajima et al.\ (2006) interprets the luminosity dependence 
of the resonance energy in terms of the correlation between the X-ray luminosity 
and the column height 
assuming the dipole shape of the magnetic field\cite{nakajima06}. 
The studies are closely related to the basic questions 
on the accretion manner along the magnetic field line 
and/or the magnetic-field structures of neutron stars.

Among 15 X-ray pulsars showing CRSFs,
several objects show harmonic feature of the resonance absorption lines. 
One of the most impressive demonstrations is the detection of 
four harmonics from X0115+634 with {\it BeppoSAX} \cite{Santangelo99}. 
The resonance energy of higher harmonics is strongly related 
to the quantum electromagnetism; 
whether the ratio of resonance energies between the fundamental 
and second harmonics is exactly 2.0 or not. 
Currently, it is consistent with 2.0 within large statistical errors 
with the recent measurements, 
and thus more sensitive instruments in the harder X-ray band are needed. 
Furthermore, recently, {\it RXTE} found a hint 
of luminosity dependence of the ratio of resonance energies, 
posing several interesting questions 
on the radiation transfer in the accretion column 
under the strong magnetic field \cite{Nakajima10}. 

The CRSFs from many X-ray pulsars have been detected 
only in absorption. 
Recently, a Japanese group reported a hint of 
a cyclotron resonance 'emission' feature instead of absorption 
from a neutron star 4U1626--67 in the dim phase with {\it Suzaku}
\cite{Iwakiri10}. 
Considering that the atmospheres on magnetized white dwarfs 
are known to show cyclotron emission features in the ultraviolet band
\cite{Schwope90}, 
whether the feature is in absorption or emission 
may depends on the optical depth of the plasma; 
i.e., optically thin plasma produces emission lines, whereas
thick case produces absorptions.
In the case of 4U1626--67, 
the emission feature was found only in a dim phase and 
other phases show absorption spectra, 
thus, the variation of the optical depth by its directions is suggested.

\subsubsection{Future prospects}
\label{subsubsection:science:plasma:future}

With future missions, we will get wide-band spectroscopic capabilities 
and polarization instruments. 
From the recent observations, 
we can set the following fundamental questions to solve with future missions.
\begin{itemize}
\item Geometry of plasma and magnetic field on neutron stars, 
      whether it is dipole field or not
\item Cyclotron resonance physics related to quantum magnetism
\item Radiation transfer in the accretion plasmas under strong magnetic field
\end{itemize}

To solve the first item, 
we need to search for CRSFs from X-ray pulsars 
in lower X-ray luminosity phase. 
Several X-ray pulsars show 
the luminosity dependence of the resonance energies, 
but some other objects do not, 
even when their luminosities are two orders of magnitude lower 
than those in the brighter phase \cite{Terada06}, 
where the fluxes are comparable to the sensitivity limit 
of the current instruments like HXD on-board {\it Suzaku}. 
The HXI and SGD instruments on-board {\it ASTRO-H}, 
having one or two orders of magnitude higher sensitivities 
in the hard X-ray band, will be indispensable for the study.

The second item can be recognized into two sub topics; 
resonance energies of harmonics, 
as described in section \ref{subsubsection:science:plasma:recent}, 
and the shape of resonance feature (whether it is Gaussian or Lorentz shape), 
which will help our understandings on 
the electron energy-distributions and cyclotron resonance process
\cite{Enoto08}. These two topics will request us the wide-band spectroscopy, 
which will be achieved soon by the future mission, {\it ASTRO-H}.

\begin{figure}[htb]
\sidecaption
\centerline{
\includegraphics[width=0.45\columnwidth,angle=-90]{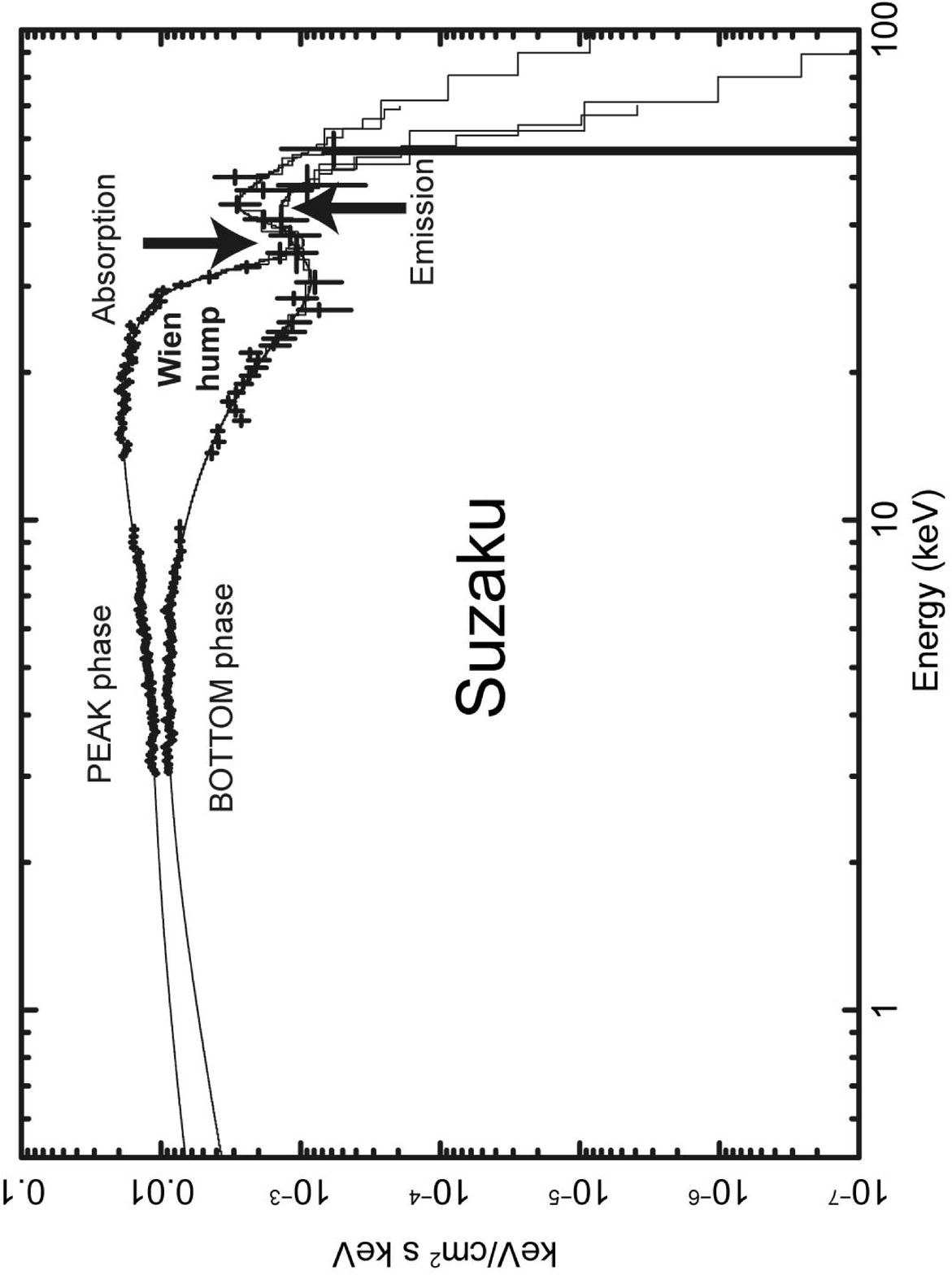}
\includegraphics[width=0.45\columnwidth,angle=-90]{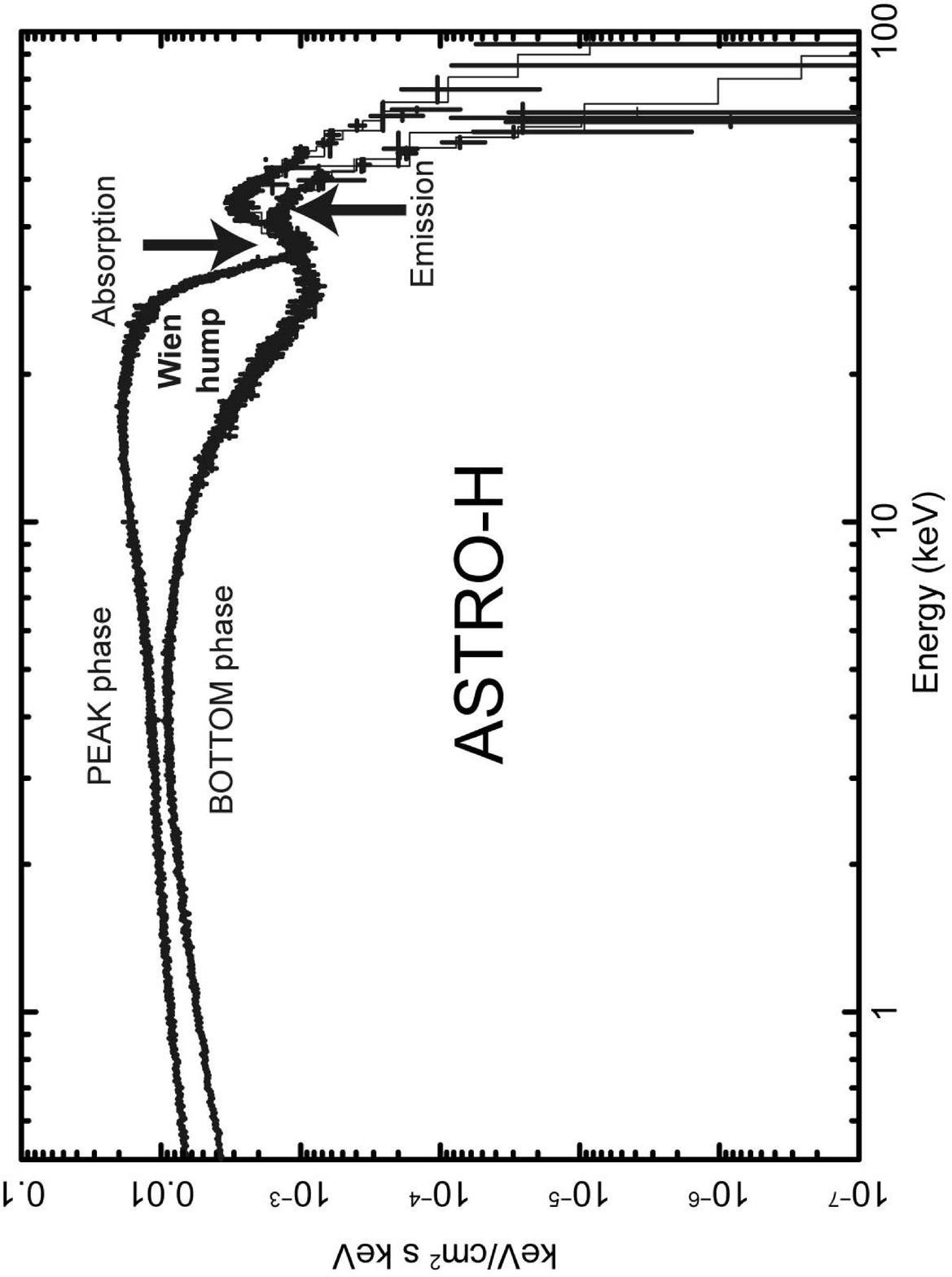}
}
\caption{(left) X-ray spectra of 4U1626-67 at bright and dim phases 
of the spin modulation taken with {\it Suzaku}. The best fit models
are shown in line. (right) The same X-ray spectra but expected for
the {\it ASTRO-H} observation with 100 ksec exposures.}
\label{fig:science:plasma:4u1626}
\end{figure}

To have new information on the third topic,
as well as numerical approaches on the radiation transfer 
in accretion columns, we have to try first, observationally,  
to verify the cyclotron resonance emission feature, 
which was statistically insignificant with the current mission. 
As demonstrated in Fig.\ref{fig:science:plasma:4u1626}, 
{\it ASTRO-H} will give us fine phase-resolved
X-ray spectra with higher qualities 
to distinguish it is absorption or emission. 
As already described in section \ref{subsubsection:science:plasma:recent},
the emission may be caused by an anisotropy of the optical depth 
of cyclotron resonance photons in plasmas of neutron stars.
This should be an application of higher optical-depth cases
of accretion plasmas of cataclysmic variables,
where we can observe anisotropic transfer of atomic-resonance photons 
in plasmas on the magnetic pole of white dwarfs \cite{Terada01, Terada04}.
In addition, polarization measurements around the resonance energy 
with future missions, like {\it ASTRO-H} and {\it IXO} 
({\it GEMS} does not cover the hard X-ray band), 
will give us more information on the transfer of 
X-mode and O-mode photons under strong magnetic field \cite{meszaros79}.

\subsection{Emission mechanism from Magnetars}
\label{subsection:science:magnetar}
Anomalous X-ray Pulsars (AXPs) and Soft Gamma-ray Repeaters (SGRs) 
are recognized as neutron stars 
with super strong magnetic fields reaching $10^{15}$ Gauss, 
named magnetars \cite{duncan92}. 
Mysterious features of magnetars still pose us many fundamental questions.
Why they have so strong magnetic field? 
What is the energy source of the X-ray emissions? 
How they are formed?

\subsubsection{Recent Studies}
\label{subsection:science:magnetar:recent}

Pulsars are rotating magnets, 
and their electromagnetic radiations are normally originated from
the rotational energy of the neutron star. 
Actually, X-ray emission from normal pulsars are only 0.1 -- 1 \% 
of the spin-down energy loss \cite{possenti02}. 
One of most mysterious feature of magnetars is their X-ray luminosities, 
which always exceed the rotational energy loss. 
Thus, the energy source of the radiation of magnetars 
is a mysterious question. 
One possible answer is a decay of magnetic field.
Pons et al (2007) found a clear relation 
between the magnetic field strength 
and the effective temperature of the neutron star surface 
of normal pulsars and magnetars, 
and they argue that the decay of magnetic field heats 
the surface of neutron stars especially for magnetars \cite{pons07}. 
Recently, theoretical groups proposed a conversion mechanism 
from magnetic energy into radiation \cite{Cooper10}.
However, the energy source of magnetars emissions 
and its conversion mechanism are still open question.

Another important discovery of magnetars is 
the hard X-ray component above 10~keV, 
from 1E 1841--045 and other three AXPs 
with {\it INTEGRAL} and {\it RXTE} \cite{kuiper04, kuiper06}. 
Recently, in addition to the flare phases, 
such a hard tail was fond even in the quiescent phase 
from a magnetar 1E 1547.0--5408 with {\it Suzaku} \cite{enoto10}. 
Since the components have a very hard photon index of about 1.0 and 
have a large pulsed fraction, 
the radiation energies will exceed the spin-down energies even largely
if the hard X-ray emission is included.
The nature of this component is also an open question.

In addition to the above two questions 
related to the fundamental physics on magnetic field of magnetars, 
another importance in the astrophysics is the evolution of magnetars. 
'How they are formed? '
Hints on this question can be found in 
a recent discovery of a pulsar wind nebula near the magnetars, 
1E1547.0-5408  \cite{jacco09}, 
and an important report of the association of two magnetars candidates 
with a TeV supernova remnant, CTB37B \cite{Halpern10}.

\subsubsection{Future prospects}
\label{subsection:science:magnetar:future}

We have many fundamental questions on magnetars. 
As presented in the previous section \ref{subsection:science:magnetar:recent},
we can pick up the following three topics;
\begin{itemize}
\item What is the energy source of radiation?
\item What is the origin of the hard X-ray component?
\item How the magnetars are formed?
\end{itemize}

The first step for future missions to study the first question is 
to perform the direct measurement of the magnetic field strength of magnetars. 
Do they really have a magnetic field reaching $10^{14-15}$ Gauss? 
This value is indirectly estimated from the spin frequency and
its derivative
assuming that the energy loss is completely due to a dipole magnetic radiation.
However, there is no observational justifications for this assumption,
because no electromagnetic emission originated from the dipole magnetic
radiation is detected.
Thus, it is important to measure the magnetic field strength directly 
as a first step. 

In future missions, we have two ways to measure magnetic field strength 
directly. 
One method is to search for a possible proton cyclotron resonance feature 
in the X-ray spectra of magnetars 
like electron CRSFs shown in section \ref{subsection:science:plasma}, 
since the proton resonance energy will come into the soft energy range 
when the magnetic field strength is in the range of $10^{14-15}$ Gauss.
Observationally, verification of the absorption feature 
as the proton cyclotron line is critical for this study,
which can be achieved by detecting higher harmonics features 
in addition to the fundamental line. 
Therefore, high-energy resolution spectroscopy 
with {\it ASTRO-H} and {\it IXO} will help us. 
The other method is to perform polarization measurements 
in the soft energy range. 
The polarization fraction between the linear and circular polarizations, 
which will be available with {\it GEMS} and {\it IXO}, 
will give us information 
to constrain the magnetic field strength of magnetars \cite{Adelsberg09}. 

To study the second topic, 
the wide-band spectroscopy in the sub MeV band is a discovery space 
of the hard X-ray component of magnetars,
in order to distinguish many kinds of emission models, like 
Synchrotron emission, non-thermal bremsstlahlung, 
super thermal bremsstrahlung, 
or a broad emission via photon splitting effect 
under a quantum critical magnetic field. 
In addition, to promote the third topic, 
continuous efforts of hard X-ray surveys of magnetars 
and X-ray follow-up observations of TeV sky are important. 
Hard X-ray imaging capability with {\it NuSTAR} and {\it ASTRO-H} 
will be very helpful.

\subsection{Diversity of Pulsar systems: white dwarf pulsars}
\label{subsection:science:wd}

\begin{figure}[hbt]
\sidecaption
\centerline{
\includegraphics[width=0.48\columnwidth]{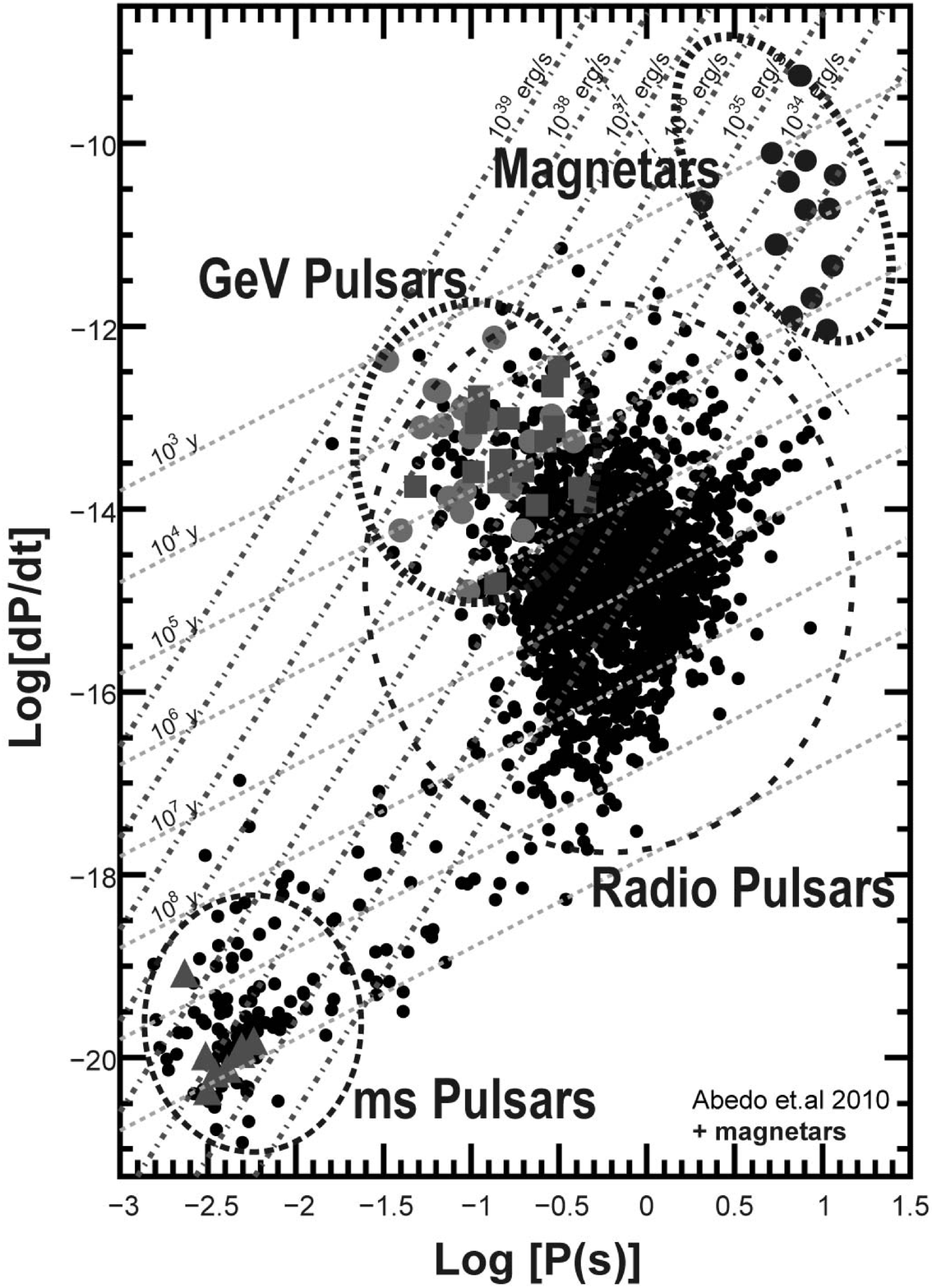}}
\caption{Scatter plot between period and period derivative of
neutron stars \cite{fermipsr10}.}
\label{fig:science:wd:ppdot}
\end{figure}
The final topic is the diversity of pulsar systems. 
Basic parameters in pulsar systems should be only three; 
magnetic field strength, characteristic age, and the inclination
of the magnetic dipole axis to the spin axis.
However, we have so many kinds of systems like radio pulsars, 
millisecond pulsars, magnetars, 
CCOs, XDINS, GeV pulsars, TeV nebulae, etc, 
as plotted in Fig.\ref{fig:science:wd:ppdot}. 
Why so highly diverse? 
What is the unified picture of pulsars? 
These questions are one of the final goals of pulsar physics.

\subsubsection{Recent Studies}
\label{subsection:science:wd:recent}
To find out hints on the question of the large diversity of pulsar systems, 
one interesting approach is to search for white dwarf pulsars.

\begin{figure}[htb]
\sidecaption
\centerline{
\includegraphics[width=0.5\columnwidth]{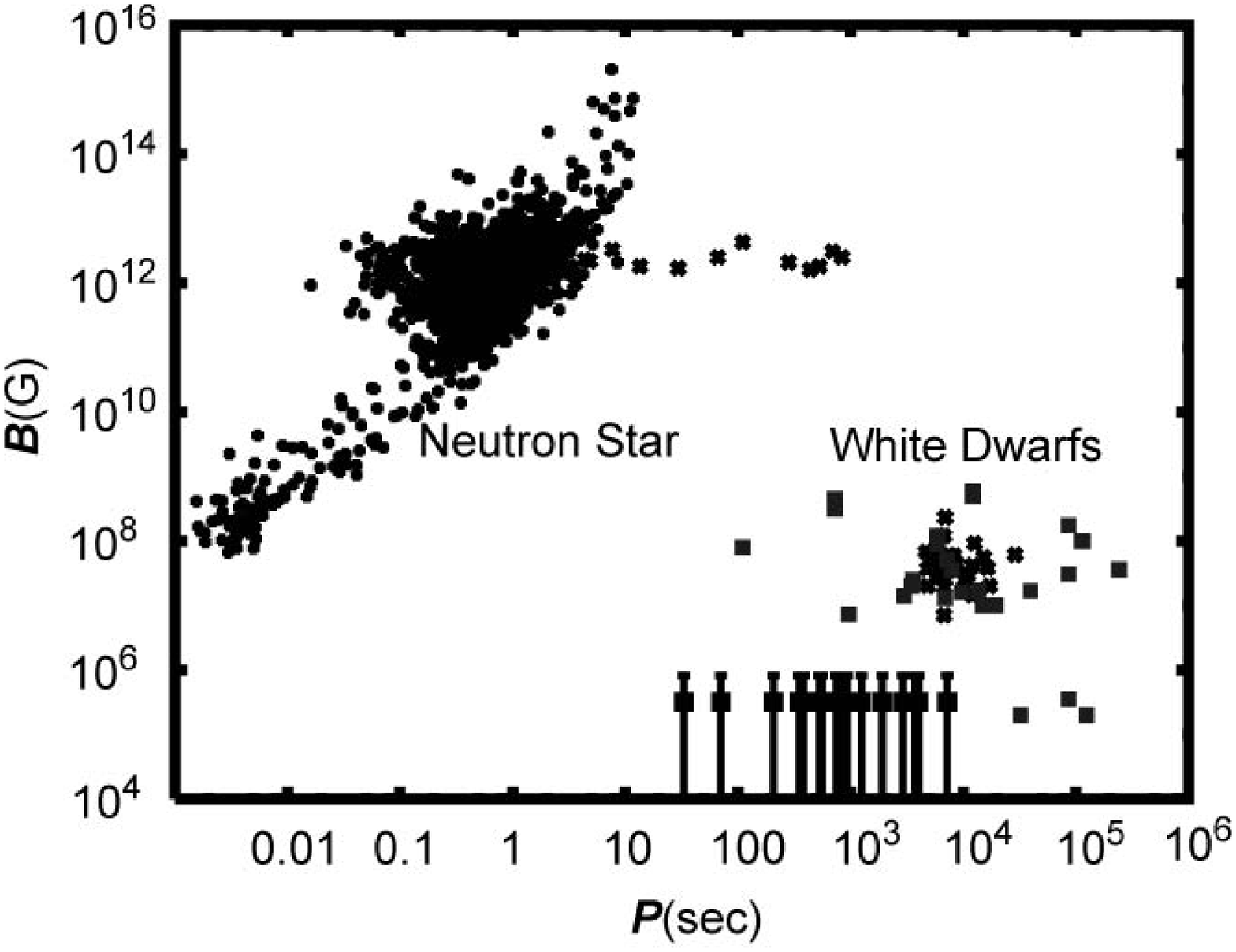}
\includegraphics[width=0.5\columnwidth]{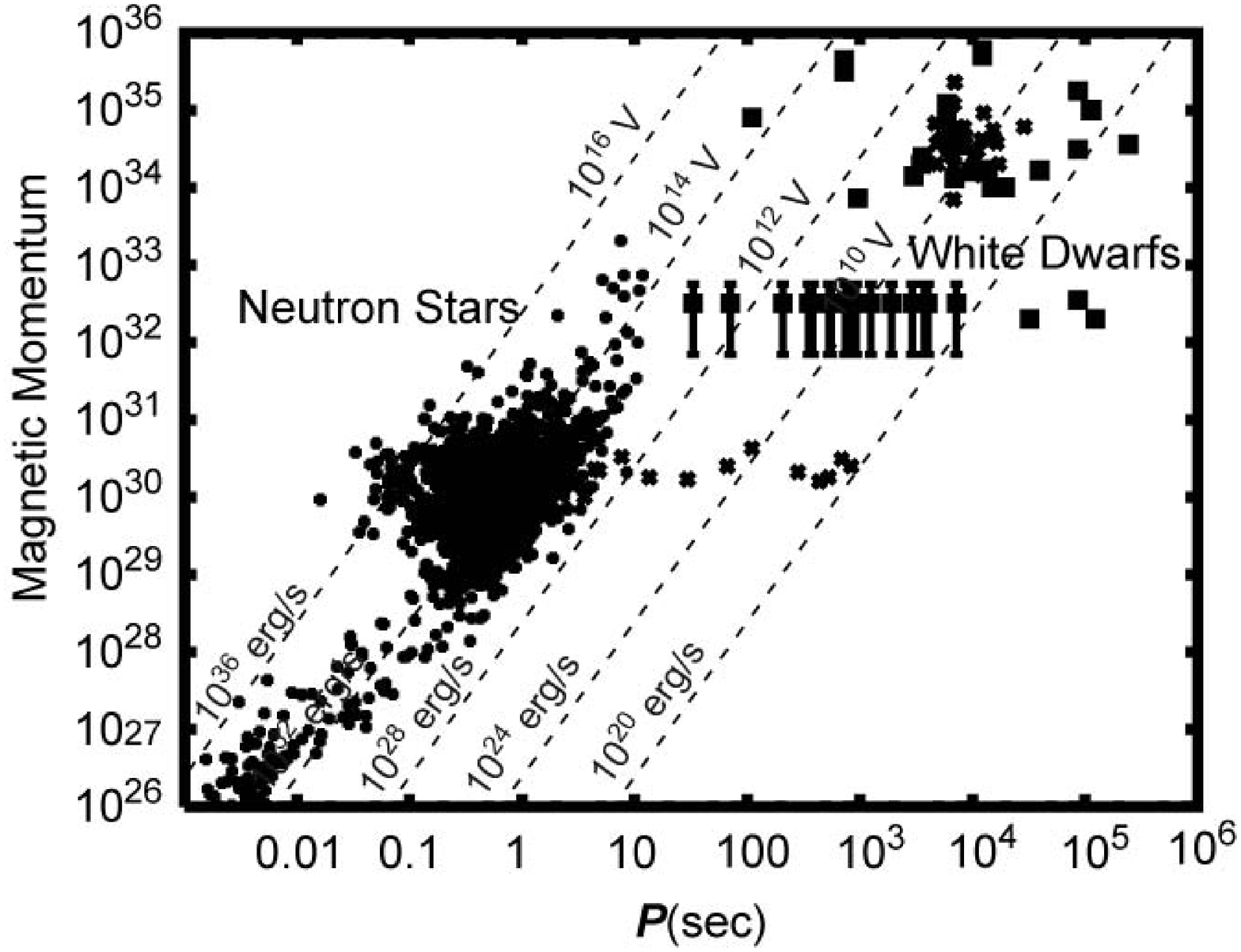}
}
\caption{Scatter plot between period and magnetic field strength (left)
or magnetic moment (right) of neutron stars and white dwarfs 
\cite{PSR1, makishima99, WD1, Terada04, kauwa06}.}
\label{fig:science:wd:pb}
\end{figure}

Why we do not have WD pulsars? 

Magnetosphere of neutron stars and white dwarfs should be the same; 
basically, they are the rotating magnets.
Generally speaking, white dwarfs have weaker magnetic field
and longer spin periods as shown in Fig.\ref{fig:science:wd:pb} left.
Normally, it is hard to observe neutron stars with 
long spin period below the death line.
Are emission mechanisms in rotating compact objects,
like rotation-powered pulsars, turned off in white dwarf systems?
To compare the systems between neutron stars and white dwarfs,
we can make a scatter plot of the magnetic momentums,
instead of surface magnetic fields,
as shown in Fig.\ref{fig:science:wd:pb} right. 
So the several magnetized white dwarfs are the kind of magnetars. 
Observationally, non-thermal incoherent radio emission has in fact 
been detected from seven WDs via systematic radio surveys
\cite{mcv_radio1,mcv_radio2,mcv_radio3,mcv_radio4,mcv_radio5},
and several objects were reported to show TeV emission
\cite{aeaqr_tev1, aeaqr_tev2,amher_tev}.

If we find out white dwarf pulsars, 
we can survey wide parameter ranges of the magnetic fields, 
the spin periods, and the system scales, 
by two or three orders of magnitude from the neutron star cases. 
In addition, if they generate high energy particles, 
they should be very important on galactic cosmic-ray origins, 
because white dwarfs exist everywhere in the universe.

\begin{figure}[htb]
\sidecaption
\centerline{
\includegraphics[width=0.45\columnwidth]{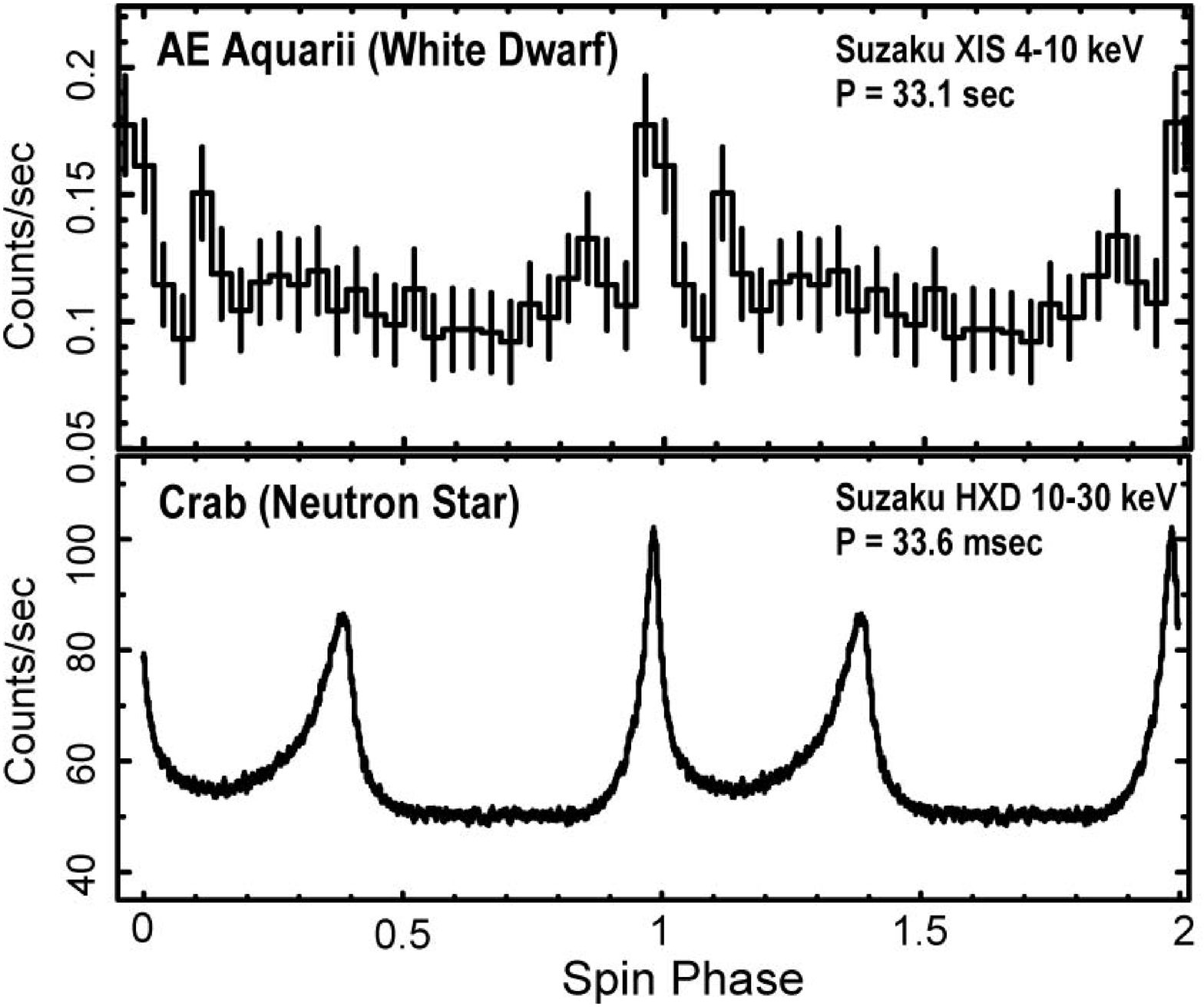}
\includegraphics[width=0.45\columnwidth]{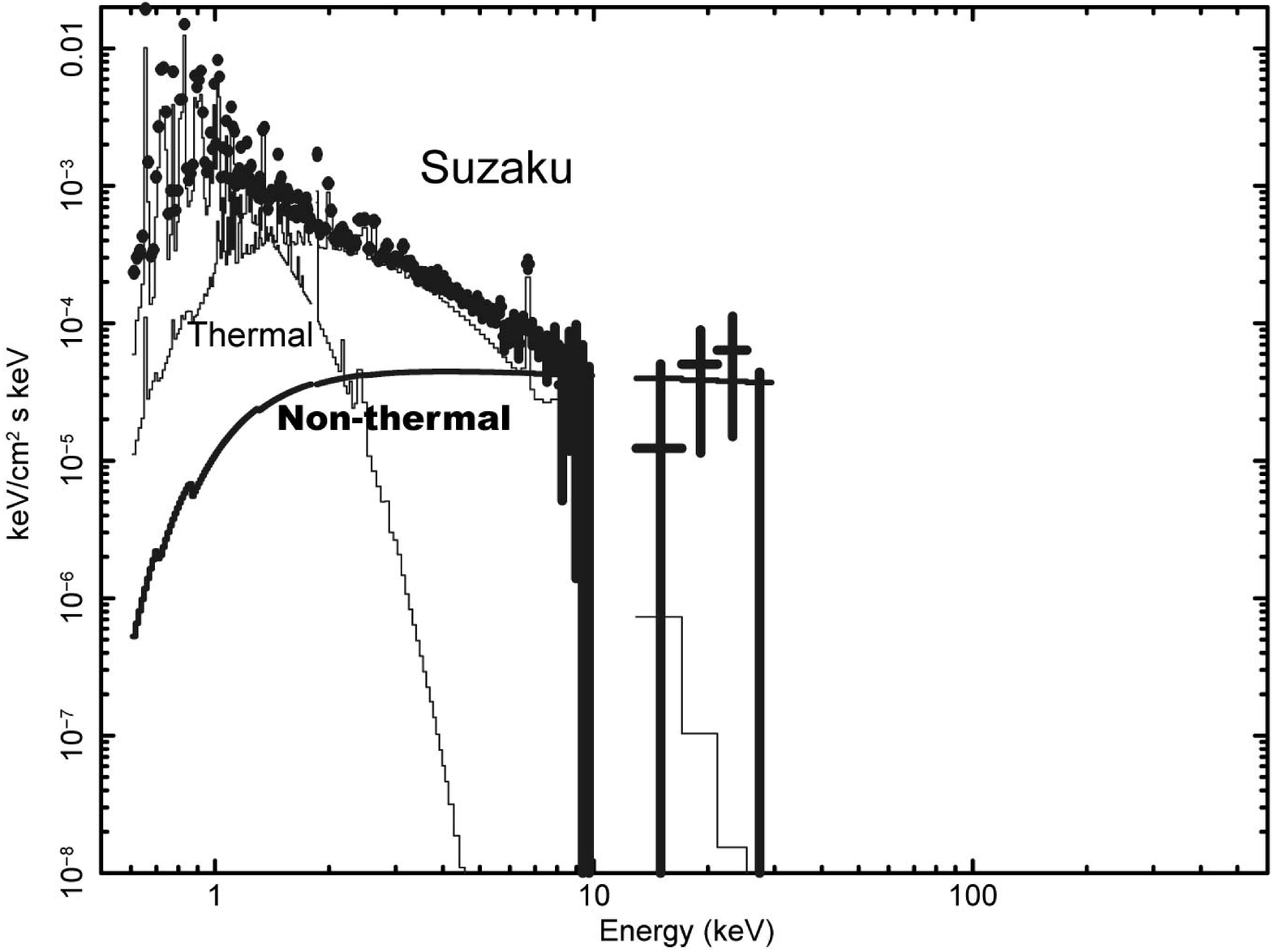}}
\caption{Left panel show the spin profile in X-ray band 
of AE Aqr and Crab pulsar taken with {\it Suzaku} 
\cite{Terada08a, Terada08b}. Right panel shows 
the X-ray spectrum of AE Aqr with {\it Suzaku} \cite{Terada08b}.}
\label{fig:science:wd:aeaqr}
\end{figure}

Among the magnetized white dwarfs, 
AE Aqr is one promising object of white dwarf pulsar 
\cite{deJager94, Ikhsanov99}, 
and recently the object is observed in the X-ray band 
with a purpose of searching for possible non-thermal emission 
with {\it Suzaku} \cite{Terada08b}. 
They find some marginal pulsation like neutron stars 
as shown in Fig.\ref{fig:science:wd:aeaqr} left. 
With the current mission, 
the X-ray spectrum can be interpreted either
by thermal component with very high temperature 
or a non-thermal power-law model,
as shown in Fig.\ref{fig:science:wd:aeaqr} right.
The X-ray luminosity of the hard X-ray component 
comes to about 0.5\% of the spin down luminosity of the object; 
this number is quite similar to the neutron star cases \cite{possenti02}.
Therefore, the AE Aqr is most promising candidate of the white dwarf pulsar.

\subsubsection{Future prospects}
\label{subsection:science:wd:future}

\begin{figure}[htb]
\sidecaption
\centerline{
\includegraphics[width=0.31\columnwidth,angle=-90]{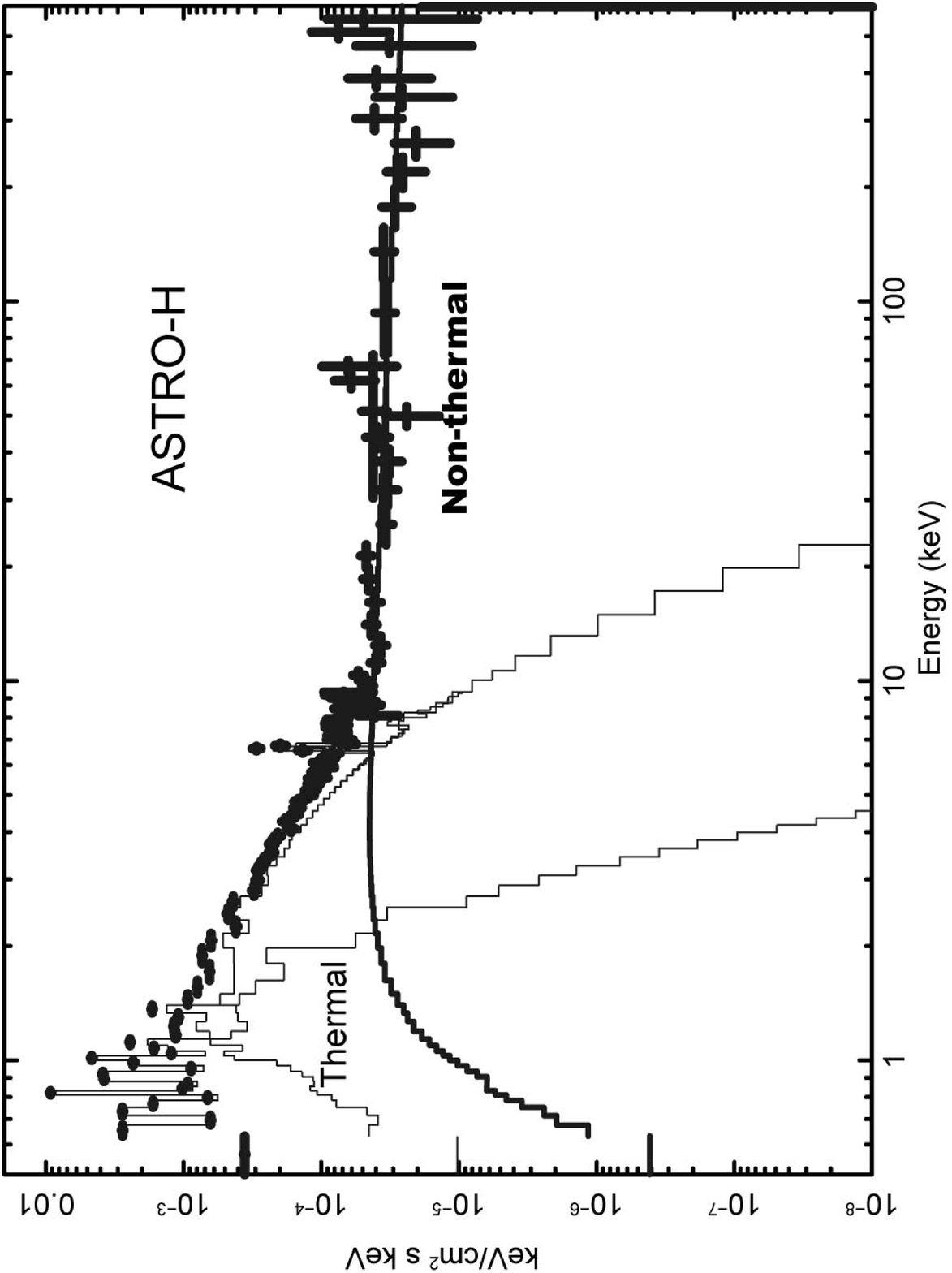}
\includegraphics[width=0.31\columnwidth,angle=-90]{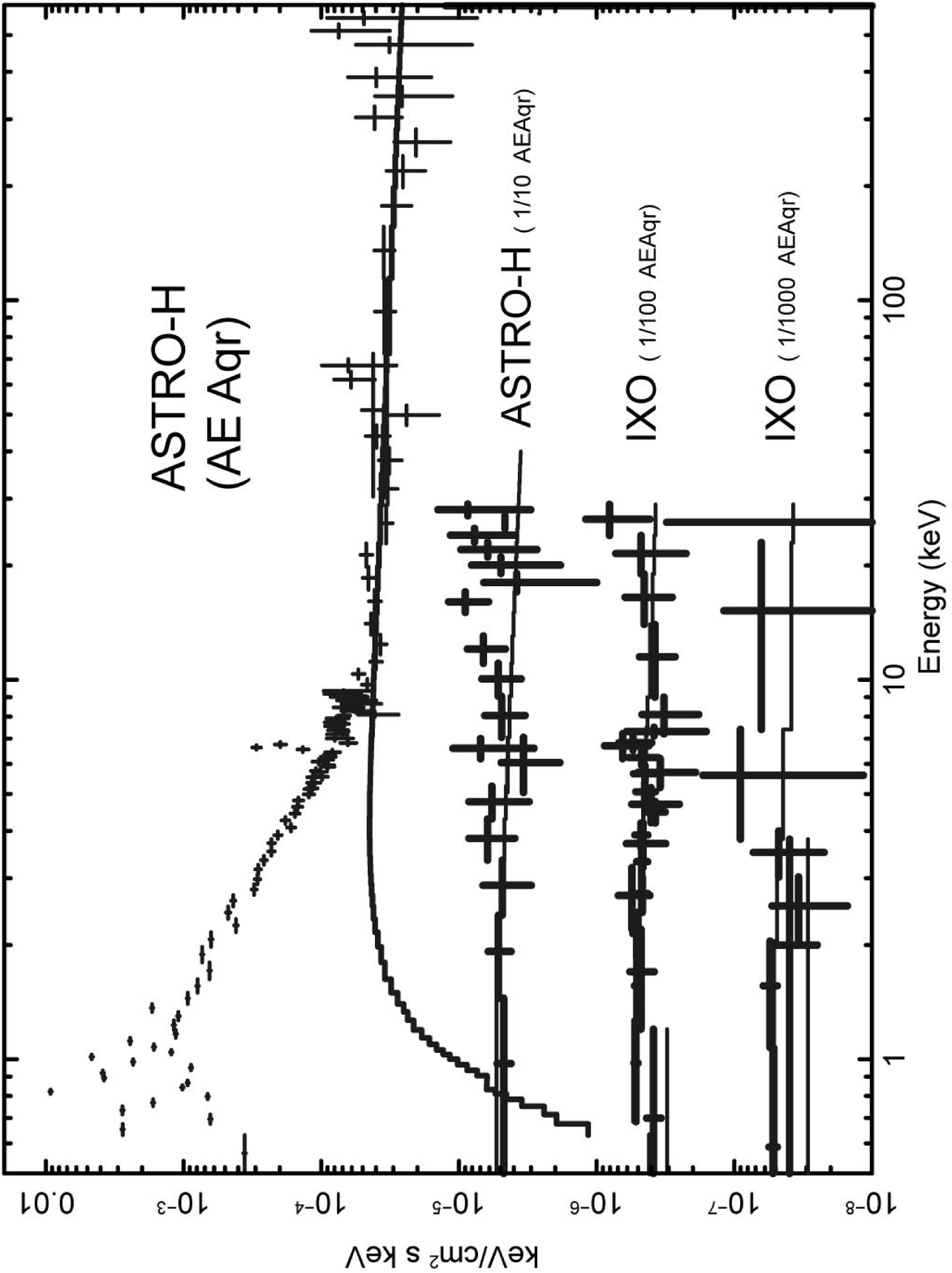}
}
\caption{Left panel show the X-ray spectrum of AE Aqr 
expected for the {\it ASTRO-H} mission with 100 ksec exposure. 
Right panel show X-ray spectra of possible non-thermal emission
from objects with the X-ray flux of 1/10, 1/100, 1/1000 of AE Aqr 
with {\it ASTRO-H} or {\it IXO} with 100 ksec exposures.}
\label{fig:science:wd:spectra}
\end{figure}

In the future, first of all, 
we have to try to confirm the first white dwarf pulsar candidate, AE Aqr. 
Normally, the X-ray emissions from cataclysmic variables 
show thermal radiations in the soft X-ray band below 20 keV, 
and thus the contamination of thermal emissions will easily 
disturb the survey of possible non-thermal emission 
behind the bright thermal radiation.
However, {\it ASTRO-H} will detect non-thermal component 
in the hard X-ray band clearly, 
as demonstrated in Fig.\ref{fig:science:wd:spectra} left.

In addition, we have more samples in isolated magnetized white dwarfs,
from which thermal emissions from accretion plasmas are expected to
be smaller than those in cataclysmic variables.
With super high sensitivities of {\it IXO},
we can survey about three orders of magnitude dimmer objects than AE Aqr
as shown in Fig.\ref{fig:science:wd:spectra}; 
quantitatively, about 30 samples are there in the current catalog. 
The discovery space of white dwarf samples is now expanding 
with the deep optical survey, SDSS \cite{sdss06}. 
Finally, we will have more objects 
to compare characteristics between white dwarf pulsars 
and neutron star pulsars.

\section{Synagy with other wavelength observatories}
\label{section:synagy}
\begin{figure}[hbt]
\sidecaption
\centerline{\includegraphics[width=0.9\columnwidth]{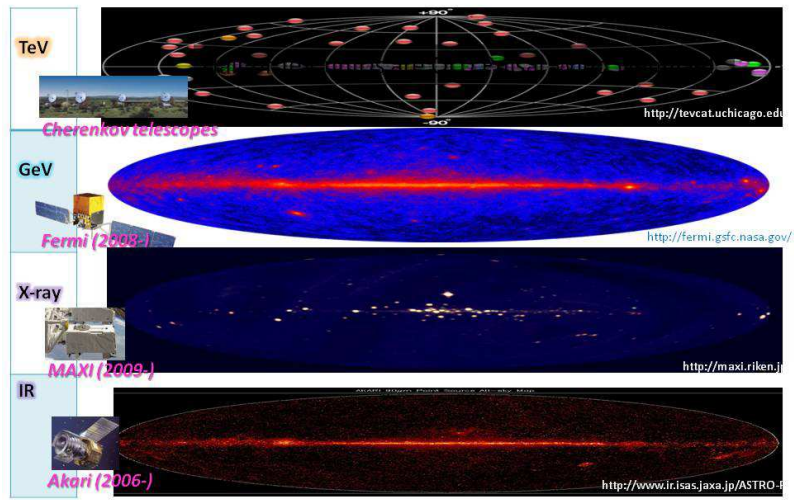}}
\caption{The all sky maps in TeV, GeV, X-ray, 
and infrared energy bands, from top to bottom, respectively. }
\label{fig:synagy1}
\end{figure}

Many future missions like {\it ASTRO-H} and {\it IXO} 
are pointing satellite with a narrow field of view 
but superb sensitivities. 
How can we get new interesting transient objects, 
like magnetars, accretion powered pulsars, 
and new astrophysical objects? 
Tight collaboration with instruments 
having all-sky monitoring function, shown in Fig.\ref{fig:synagy1}, 
is important for current and future missions. 

\begin{figure}[htb]
\sidecaption
\centerline{\includegraphics[width=0.9\columnwidth]{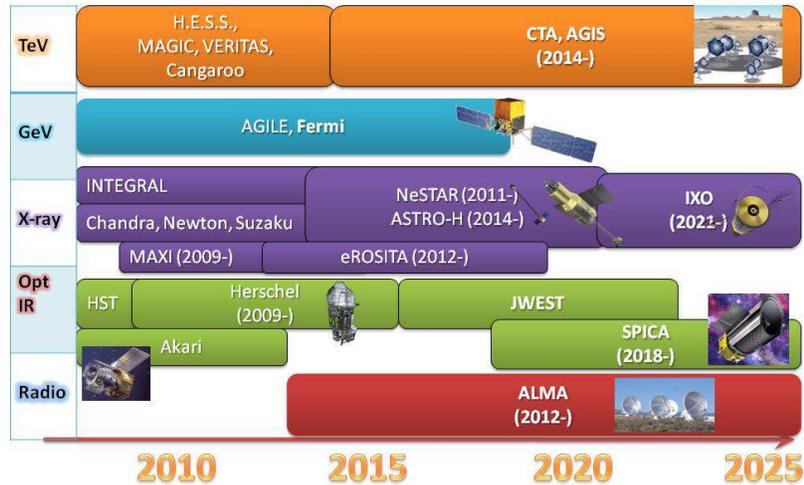}}
\caption{Near future missions in various energy bands.}
\label{fig:synagy2}
\end{figure}

For the successful X-ray observations in future, 
collaboration with other wave-length is very important. 
We live in so lucky era to have very nice missions, 
like Cherenkov Telescope Array in TeV band, {\it IXO}, 
{\it SPICA} in the infra-red band, and 
ALMA in the radio band, as illustrated in Fig.\ref{fig:synagy2}.

\end{document}